# DESIGN AND DEVELOPMENT OF AN ULTRASONIC MOTION DETECTOR

Adamu Murtala Zungeru

School of Electrical and Electronic Engineering, University of Nottingham, Jalan Broga, 43500 Semenyih, Selangor Darul Ehsan, Malaysia
adamuzungeru@ieee.org

## ABSTRACT

*The ultrasonic motion detector devices emit ultrasonic sound energy into an area of interest (monitored area), and this further reacts to a change in the reflected energy pattern. The system uses a technique that is based on a frequency shift in reflected energy to detect a movement or change in position (motion). In this system, ultrasonic sound is transmitted from the transmitting device which is normally in the form of energy. The transmitted sound utilizes air as its medium and this travel in a wave type motion. The wave is reflected back from the surroundings in the room/hallway and the device hears a pitch characteristic of the protected environment. In this system, the wave pattern is disturbed and reflected back more quickly, thus increasing the pitch and signaling an alarm whenever motion is detected. The main contribution of this work is the design of a circuit that can sense motion through movement of anything, a low cost and portable motion detector, and the design of a circuit that can be used to trigger another circuit whether to ON or OFF depending on the circuit attached to it. Generally, the design is made to detect movement or moving object in a an enclosed area. In this work, a transmitter transducer generates a signal at a frequency of 40khz, and when the signal is blocked by any moving object, and this in turn, triggers a buzzer via a timing circuit. This system works on the principle of the signal interference by a moving body, and the system is dependent on the presence of an intruder or moving object within a monitored area. The system after design and construction was tested and found to work in accordance with specifications.*

## 1. INTRODUCTION

A motion detector is a kind of security system that uses sensing ability in the form of sensors to detect movement and and this usually triggers an alarm, or sometimes activate another circuit. However, motion detectors are normally used to protect indoor areas, in this, conditions can then be controlled more closely. Detectors for use in homes for security purpose usually detect movement in a closed space area of little feet-by-feet. Detectors for large range warehouses can protect areas with dimensions as large as 24mx37m (80ft by120ft) [1]. The motion detector is normally useful in places like museums where important assets are located. As such, motion detectors can detect break-in at vulnerable points. Such points include walls, doors windows and other openings. Special motion detectors can protect the inside of exhibit cases where items such as diamonds arc placed. Others can be focused on a narrow area of coverage, somewhat like a curtain, that projected in front of a painting to detect even the slightest touch.

Motion detector systems use a variety of methods to detect movement. Each method has advantages and disadvantages. Motion detectors can be categorized into two major types [2-3] these are namely: (1) Passive detectors, and (2) Active detectors.

Passive detectors are detectors which do not send out signals but merely receive signals, such as change in temperature, change in light intensity and so on. Most infrared detectors are passive detectors. While Active detectors are detectors which send out waves of energy and receive waves reflected back from objects. Any disturbance in the reflected waves caused by example a moving object will trigger an alarm. Microwave and ultrasonic detectors are examples of active detectors.

Man and animal or moving object produces sound. The sound is created as a result of their physical movement, which might be low or fast movement, and also depends on the medium that create the sound. However, these movements can be detected by using an ultrasonic sensor. The ultrasonic sound waves are sound waves that are above the range of human hearing and, thus, have a frequency above about 20khz. Any frequency of above 20kz is considered ultrasonic [4-6, 8-10].

In general, an ultrasonic sensor typically comprises of one or more ultrasonic transducer which transforms electrical energy into sound and vice-versa, a casing which encloses the ultrasonic transducer, connectors, and if possible some electronic circuit for signal processing.

Nowadays there are numerous of the commercial ultrasonic motion detectors, basically the main aim of this work is to design and construct a simple and cheap ultrasonic motion detector system which is aimed at detecting the physical movement of human, animal, or anything that moves. The design is to improve the use of sensor in detecting motion. In general, it is aimed at reduction of the cost to design, develop or construct an ultrasonic motion detector.

## 2. ULTRASONIC MOTION DETECTORS

Generally, there exist numerous of motion detector, but of our interest is the ultrasonic motion detectors due to its numerous advantage over other types of detectors. For example, having fast response time and very sensitive, no physical contact required by the object, being environmentally friendly and reliable, and above all utilizing ultrasonic waves that are not visible and audible to human. Ultrasonic motion detectors are electrical devices, which use ultra-sound (that is, sound of very high frequency) to detect motion. In such a detector a transmitter emits a sound of a frequency which is normally too high for the human ear to hear. When a receiver picks up the sound waves that is reflected from the area under protection, it sends it to an appropriate circuit for further action (normally an audio circuit). In the case of motion of human or target in the space between the receiver and transmitter, further change, or shift in the frequency of sound is experienced [3, 7], a circuit in the device detects any unusual shift in frequency, which is normally noted due to predefined frequency. A small shift in frequency, such as that produced by an insect or rodent, is ignored. When a noticeable shift is observed, such as a large shift produced by a moving person, the device triggers the alarm.

### 2.1. Mode of Operation

The ultrasonic motion detector uses a phenomenon known as the "Doppler Effect" in detecting the motion of an object. The Doppler Effect is the apparent difference between the frequencies at which sound or light waves leaves a source and that at which they reach an observer, caused by relative motion of the observer and the wave source. Examples of the Doppler Effect include:

1. As one approaches a blowing horn, the perceived pitch is higher until the horn is reached and then becomes lower as the horn is passed.
2. The light from a star, observed from the earth, shifts towards the red end of the spectrum (lower frequency) if the earth and the star are receding from each other, and towards the violet (higher frequency) if they are approaching each other.

### 2.2. Doppler Shift Derivation

Consider the relationship between the frequency of sound produced by a source moving with velocity V and the frequency received by a receiver moving with velocity $V_r$. For simplicity, we assume that both the source and the receiver are moving in a straight line in the same direction. At time t = 0, the source (S), and receiver (R) are separated by a distance (d).

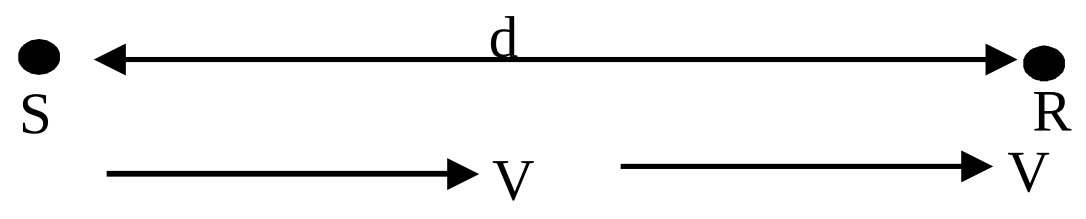

The source emits a wave that propagates at a velocity *c*, and reaches the receiver after time t as the receiver has moved $V_r t$ meters

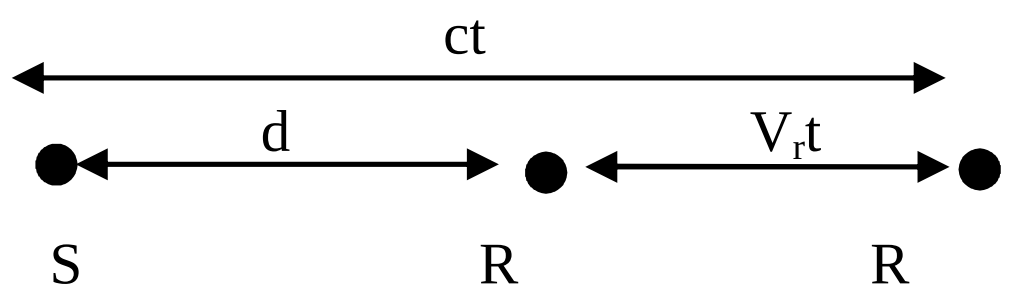

Therefore $ct = d + V_r t$ (1)

$$t = \frac{d}{c - Vr} \quad (2)$$

At time t, the source *(S)* would have moved $V_s t$ meters. Let the wave emitted at that instant be received at time t' by the receiver *(R)*, in this time the receiver would have moved $V_r t'$ meters.

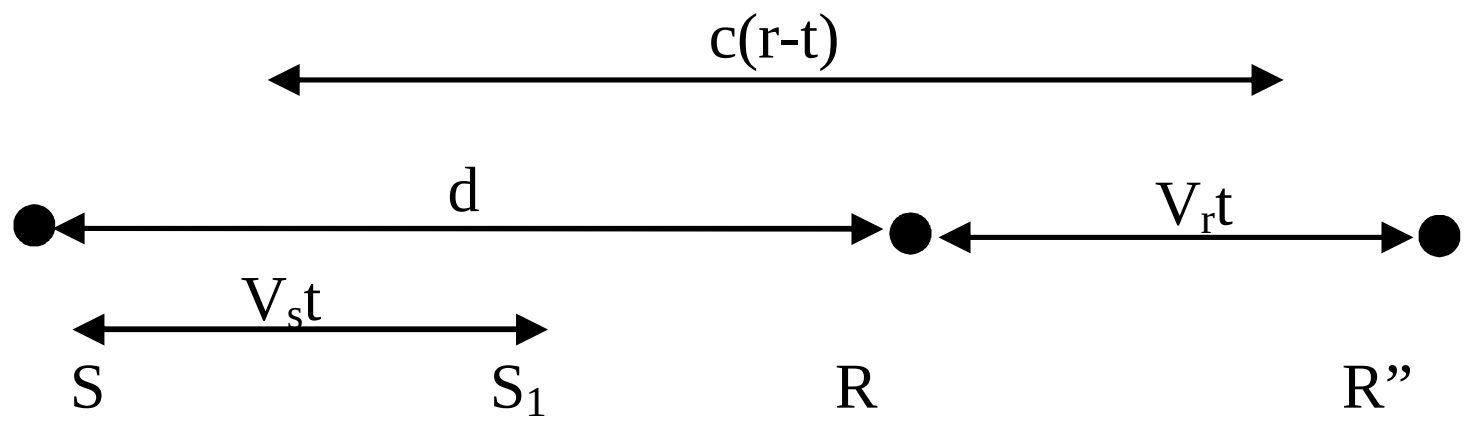

$$c(t' - t) = (d - V_s \tau) + V_r t' \quad (3)$$

Thus for the receiver, the interval between the waves has been

Hence $t' = \dfrac{d + (c - V_r)\tau}{c - V_1}$ (4)

Thus for the receiver, the interval between the waves has been

$$\tau' = t' - t = \frac{c - V_s}{c - V_r}\tau \quad (5)$$

Whereas for the source, the interval between waves has been $\tau$ .

The number of waves emitted in time *t* by the source must be equal to the number received by the receivers in $\tau'$

That is $f_r \tau' = f_s \tau$ (6)

Hence

$$f_r = \frac{c - V_r}{c - V_s} f \quad (7)$$

For $V_s$ and $V_r << c$

$$f_r = \frac{1-\frac{V_r}{c}}{1-\frac{V_s}{c}} f_s = \left[1-\frac{V_r}{c}\right]\left[1-\frac{V_s}{c}\right]^{-1} \quad (8)$$

Expanding the last term using the binomial expansion that is

$$(1+x)^n = 1+nx+\frac{n(n-1)}{2!}x^2+.... \quad (9)$$

for x << 1, the higher order terms can be ignored hence

$$f_r \approx \left[1-\frac{V_r}{c}\right]\left[1-\frac{V_r}{c}\right] = \left(1-\frac{V_{rs}}{c}\right) f_s \quad (10)$$

$$where V_{rs} = V_r - V_s \quad (11)$$

$V_{rs}$ is the velocity of the receiver relative to the source.

The Doppler frequency (shift) is thus

$$f_d = f_r - f_s = \frac{-V_{rs}}{c} f_s \quad (12)$$

The frequency moving away from the source will be less than the frequency measured at the source, whereas the frequency measured at a receiver moving towards the source will be greater than the frequency measured at the source.

## 2.3. Doppler Geometry

In most Doppler sensors both the transmitter and receiver are stationary, and they illuminate a moving target.

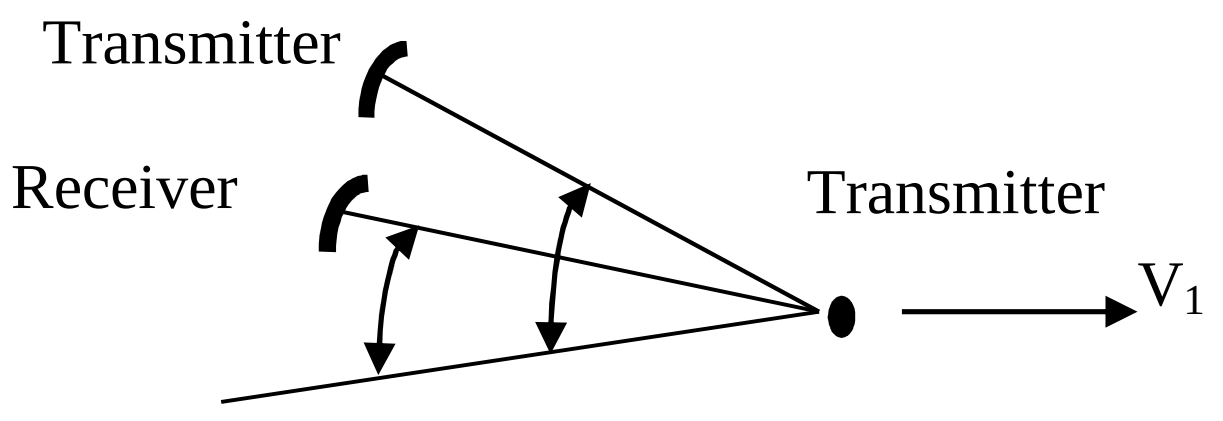

**Fig. 1:** Typical ultrasonic sensor arrangement

The velocity of the target relative to the transmitter will be $V cos \theta t$, and the velocity of the target relative to the receiver will be $V cos \theta r$. The Doppler shift arising under this circumstance can be calculated assuming that:

1. The target is a receiver moving away from the source with a velocity Vcos $\theta$ t.
2. The receiver is moving away from the target (source) with a velocity Vcos $\theta$ r.

This is equivalent to the receiver moving away from the source with a velocity $Vcos\,\theta t + Vcos\,\theta r$ even though both are stationary.

For target moving at low speeds (V<<c), the Doppler frequency for separated transducers is given as:

$$f_d = -\frac{f_s V}{c}\left(\cos\theta_t + \cos\theta_r\right) \quad (13)$$

$$f_d = -\frac{2 f_s V}{}\left(\frac{\theta_r + \theta t}{2}\right)\left(\frac{\theta_r - \theta t}{2}\right) \quad (14)$$

While for target moving at high speeds (V<c) it is not possible to use the electromagnetic radiation approximation, and the whole formula must be used for the separate transducers.

That is $f_d = \frac{c - V\cos\theta_r}{c + \cos\theta_s} f_s$ (15)

And the Doppler frequency is $f_d = f_r - f_s$ (16)

Where $f_d$ = Doppler (shift) frequency

$f_s$ = frequency at source

$f_r$ = frequency at the receiver

C = velocity of sound in air

And V= velocity of object

## 2.3. Doppler Frequency Extraction

This section is the instrumentation required to detect a Doppler shift in a received signal.

The transmitted signal is of the form $x_{r(t)} = \xi t \cos(\omega_s t)$ (17)

The correspondence received signal from a single target will be

$$x_r(t) = \xi r \cos((\omega s + \omega d)t + \theta) \quad (18)$$

Where $\theta$ = phase term dependent on the distance to the target (rad)

$\omega_s = 2\pi f_s rad / s$

$\omega_d = 2\pi_d (rad / s)$

The two signals are mixed to produce

$$x_1(t)x_s(t) = \xi_t \xi_s \cos(\omega_s t)\cos([\omega_s + \omega_s]t + \theta) \quad (19)$$

$$x_1(t)x_s(t) = \frac{\xi_r \xi_s}{2}(\cos(\omega_d t + \theta) + \cos 2((\omega_s + \omega_{d)})t + \theta) \quad (20)$$

The signal is low pass filtered to remove the component at 2*fs* leaving only the Doppler signal that is:

$$x_d(t) = \frac{\xi_r \xi_t}{2}(\cos(\omega_d t + \theta \qquad (21)$$

The reflected signal amplitude from non-moving objects in the beam will be 40dB to 50dB larger than the Doppler signal, and so additional high pass filtering is often required to remove this.

## 3. Design Analysis, Test and Measurements

The ultrasonic motion system is built around the following subsystem.

1. A 40kHz ultrasonic frequency transmitters
2. A 40kHz ultrasonic frequency receiver
3. A modulated audible alert tone generator
4. Power supply unit

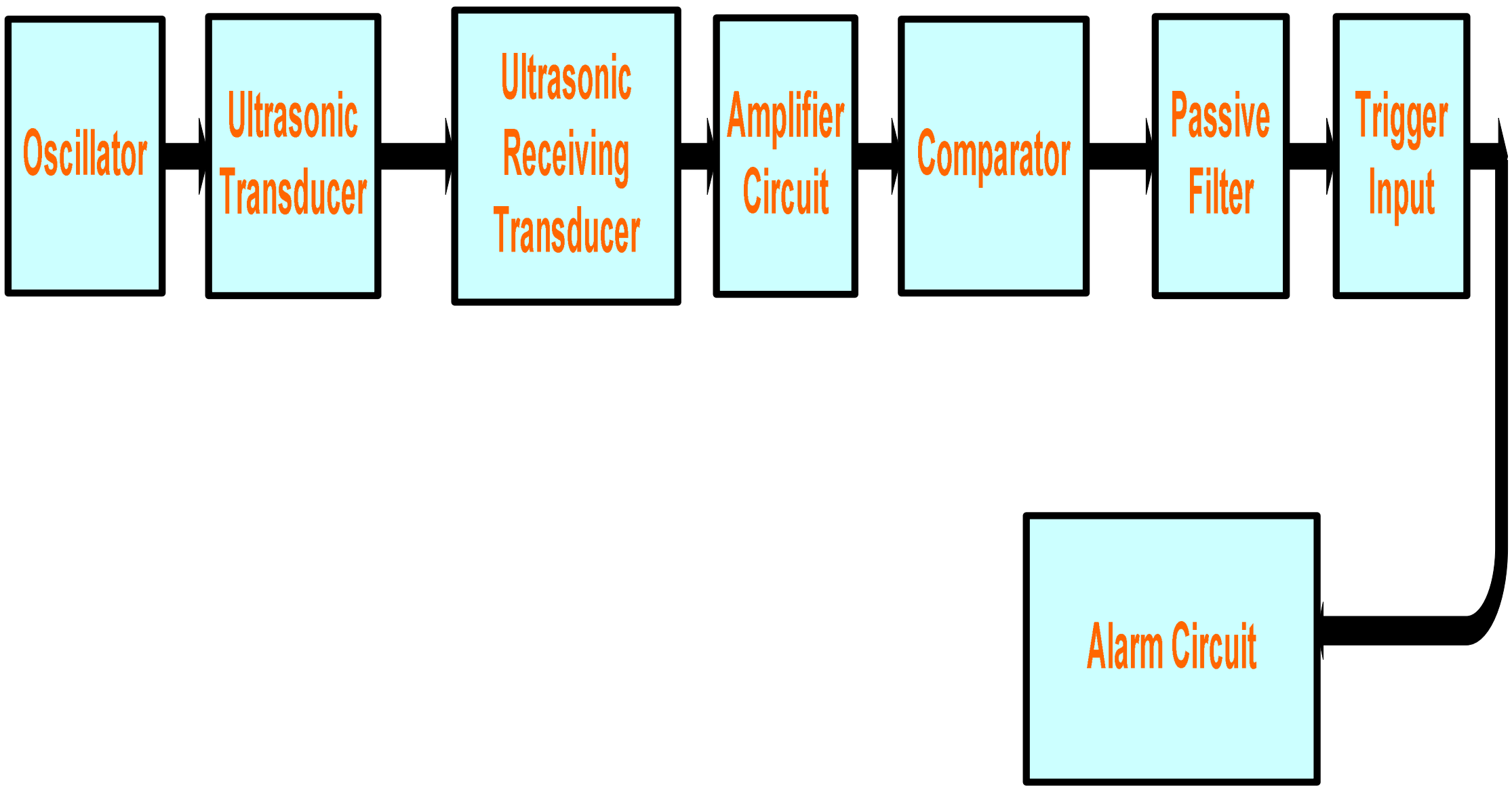

**Fig. 2**: Block Diagram for the Ultrasonic Motion Detector

### 3.1. Analysis of the 40 kHz Ultrasonic Frequency Transmitter

This section of the detector is basically built (designed) around a 555 a stable oscillator as shown in Fig. 3.

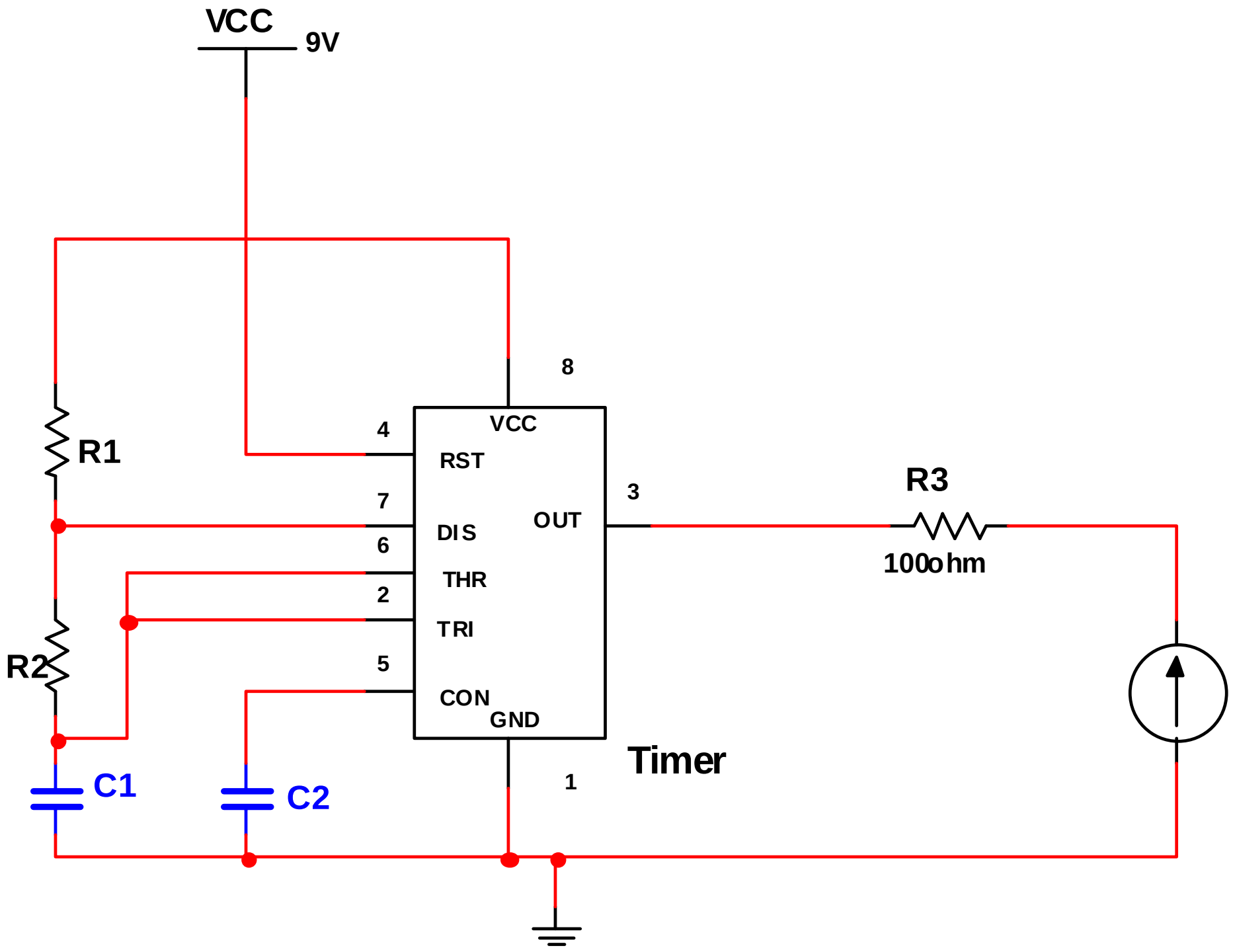

**Fig. 3**: Circuit diagram of the 40kHz ultrasonic transmitter

The astable oscillator generates a frequency of 40 kHz which is used by the ultrasonic transmitting transducer to generate the ultrasonic wave.

A stable multi-vibrator (oscillator) has no stable state consequently; it continually changes back and forth between the two states at a predetermined rate.

For clarity and better understanding we consider the complete diagram of the Astable multi-vibrator with a detailed internal diagram of the 555 timer shown in Fig. 4 below

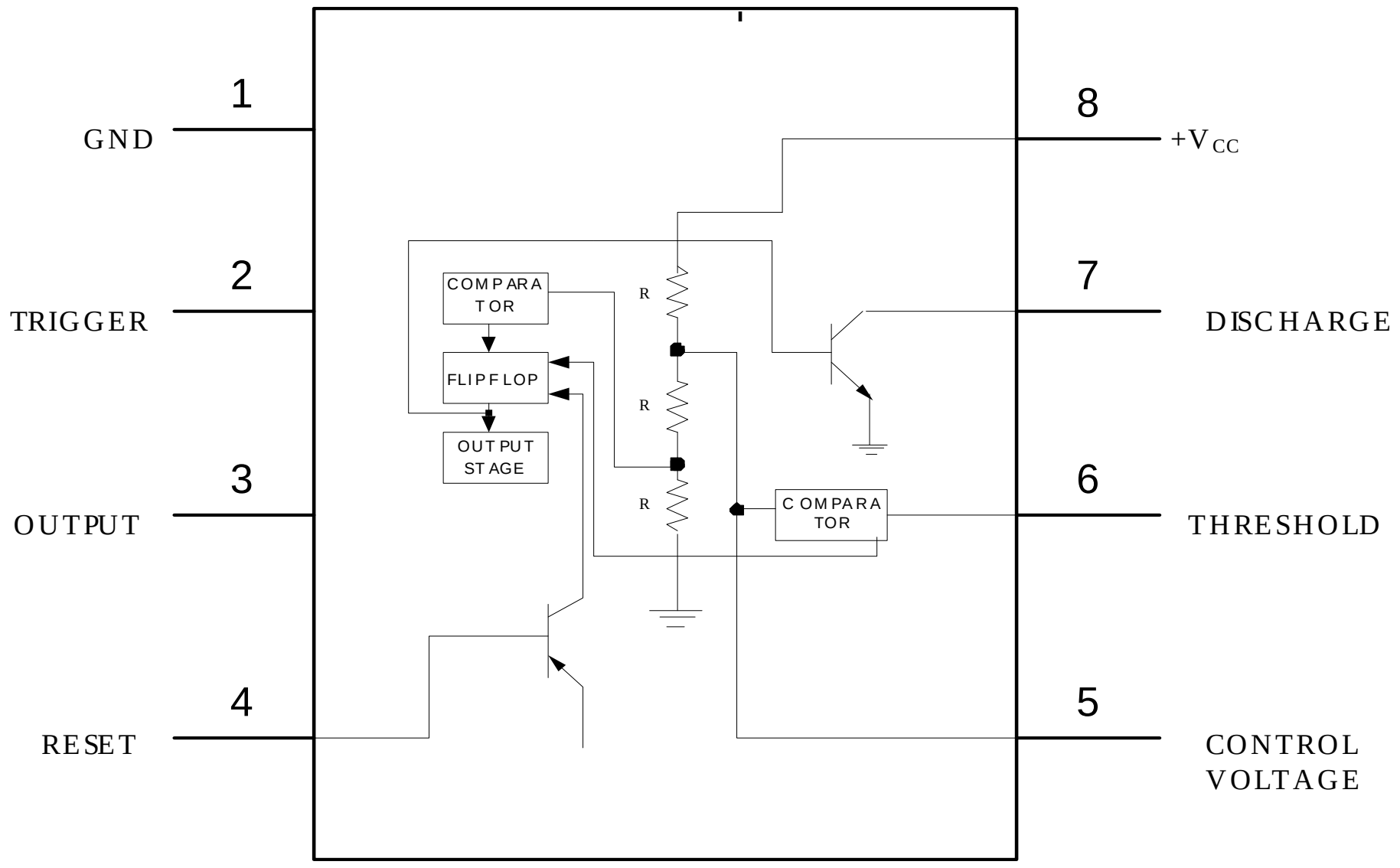

**Fig. 4:** The internal circuitry of a 555 timer

The capacitor voltage in a low pass RC circuit subjected to a step input of volts is given by

$$V_{CC} = V_{CC}[1 - e - t/RC] \quad (22)$$

The time (t1) taken by the circuit to charge from 0 to $\frac{2}{3}V_{CC}$ is given by the equation

$$\frac{2}{3}V_{CC} = V_{CC}[1-e^{-t}/RC] \quad (23)$$

Hence t1=1.09RC (24)

The time (t2) taken by circuit to charge from 0 to $\frac{1}{3}V_{CC}$ is given by the equation

$$\frac{1}{3}V_{CC} = V_{CC}[1-e^{-t}/RC] \quad (25)$$

Hence t2 = 0.405RC (26)

Hence the time to charge from $\frac{1}{3}V_{CC}$ to $\frac{2}{3}V_{CC}$ is $(t_H)$= t1-t2 (27)

Therefore $t_H$=1.09RC- 0.405RC; $t_H$= 0.6RC (28)

So the given circuit $t_H = 0.69\,(R_A+R_B)C$ (29)

The output is low while the capacitor discharge from $\frac{2}{3}V_{CC}$ to $\frac{1}{3}V_{CC}$ and the voltage across the capacitor is given by:

$$\frac{1}{3}Vcc = \frac{2}{3}Vcc\, e^{-t}/RC \quad (30)$$

Hence $t_L$=0.69RC (31)

Where $t_H$ = time taken for output to go high, and $t_L$= time taken for output to go low.

Note that both $R_A$ and $R_B$ are in the charge path, and only $R_B$ is in the discharge path hence total time (T) is given as:

$$T = (t_H + t_L) = 0.69(R_A+2R_B)C \quad (32)$$

Frequency of oscillation is given as $F = \frac{1}{T}$ (33)

### 3.2. Calculations for the 555 Timer in the Transmitter Circuit

A frequency of 40kHz is needed for this design hence from (33),

$$F = \frac{1}{T} = \frac{1.45}{(R_A + 2R_B)c} \quad (34)$$

The minimum value for $R_A$ is approximately equal to $Vcc/\,0.2A$

Therefore with supply voltage $(V_{CC})$= 9v, the minimum value for $R_A = 9/0.2 = 45\,\Omega$ hence taking the value of $R_A$= 1k$\Omega$ and C = 0.01$\mu$f and frequency of 40khz,

$$4000 = \frac{1.45}{(1+2R_B)0.01x10^{-6}}$$

Therefore $2R_B = \frac{1.45}{(1+0.01x10^{-6}x4000} - 1 = \frac{1.45}{4x10^{-4}}$

Hence $R_B$ = 1812.5k$\Omega$ $\approx$1812.5k$\Omega$ $\approx$1.8m$\Omega$

### 3.3. A 40khz Ultrasonic Frequency Receiver Control System

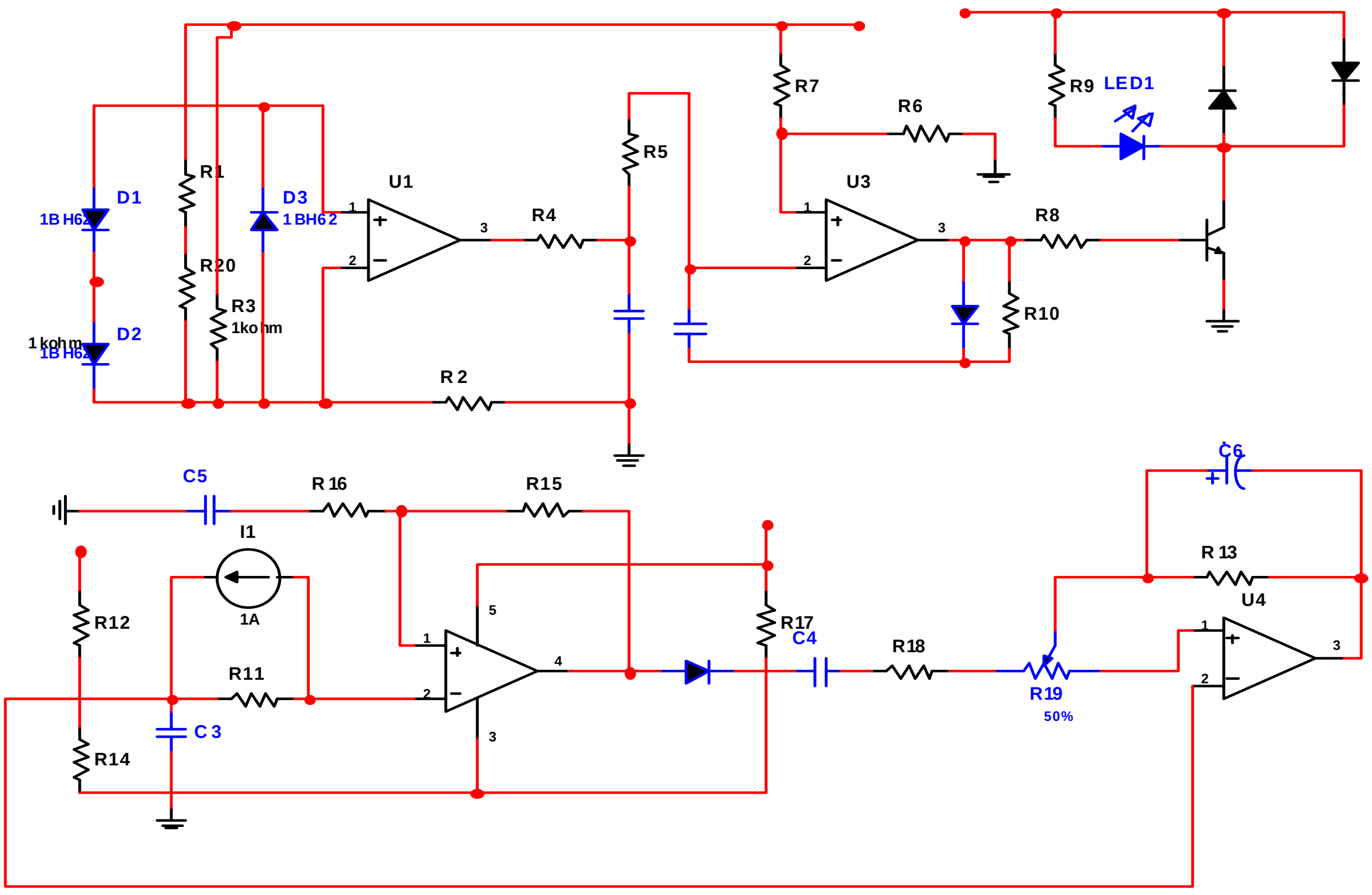

**Fig. 5:** Circuit diagram for the 40 kHz ultrasonic receiver control system

The system consist of a receiver section and is built around four AC coupled stages, in which each is further built around one of four sections of an LM 324 op-amp IC. In the first stage the input voltage developed across R1 and R2 is modulated by a 40kHz ultrasonic receiving transducer (BZ2) and is then fed to IC-la, where it is amplified. The receiving circuit consist of a transducer which detects any reflected sound produced by the transmitting transducer. In the situation when no movement is detected, the resulting envelop signal is just a straight line. If there is a movement or change in frequency, the envelope will reflect it in the form of a positive or negative signal. If there is a signal rise normally above + 0.7 volts (silicon diode breakdown voltage), part of the (D3) will conducts making the output on pin 8 to go high. If the signal falls below –0.7 volts D2 conducts, which also causes the output to go high.

The fourth stage, built ICI-d, is set up as a monostable flip-flop. Stage coverts any input pulse substantial enough to turn ON transistor Q1 conducts the LED turns on and output signals provided to drive the modulated audible tone generator.

### 3.4. Modulated Audible Alert Tone Generator

This section consists of three 555 timer ICs (that is IC2, IC3, IC4), where IC2 is operated in monostable mode, and IC3 and IC4 are in an Astable configuration. The trigger input to IOC2 (pin 2) is connected to transistor (Q2) which acts like a switch, with the base connected to the signal output of the motion detector. When the detector senses motion, even for a brief moment, a 3,5v logic is fed into the base of Q2 (fig 3.5), transistor Q3 coming ON.

Pulls pin 2 of IC2 low (ground) (thus triggering the monostable circuit. Pin 3 of IC2 then goes high. The length of time (t) it remains high is determined by the values of $R_{22}$ and C11.

That is T = 1.1 $R_{22}$ x C11. When the output of IC2 goes high, IC3 and IC4 are activated; IC4 is connected to produce a 50Hz tone, while IC3 is a 1Hz oscillator. Thus, IC3 is used to trigger IC4 ON and OFF once per second, generating a pulse–tone alarm. This is reproduced over the loudspeaker in the collector – emitter circuit of transistor Q3. The alarm sounds for as long as pin 4 of IC3 and IC4 are high when pin4 of IC2 goes low that is the monostable times out after T= 1.1 $R_{22}$ x $C_{11}$ (approximately 2 min) or the reset button is pressed, IC2 returns to its stable state then IC3, and IC4 are both reset and thus disabled.

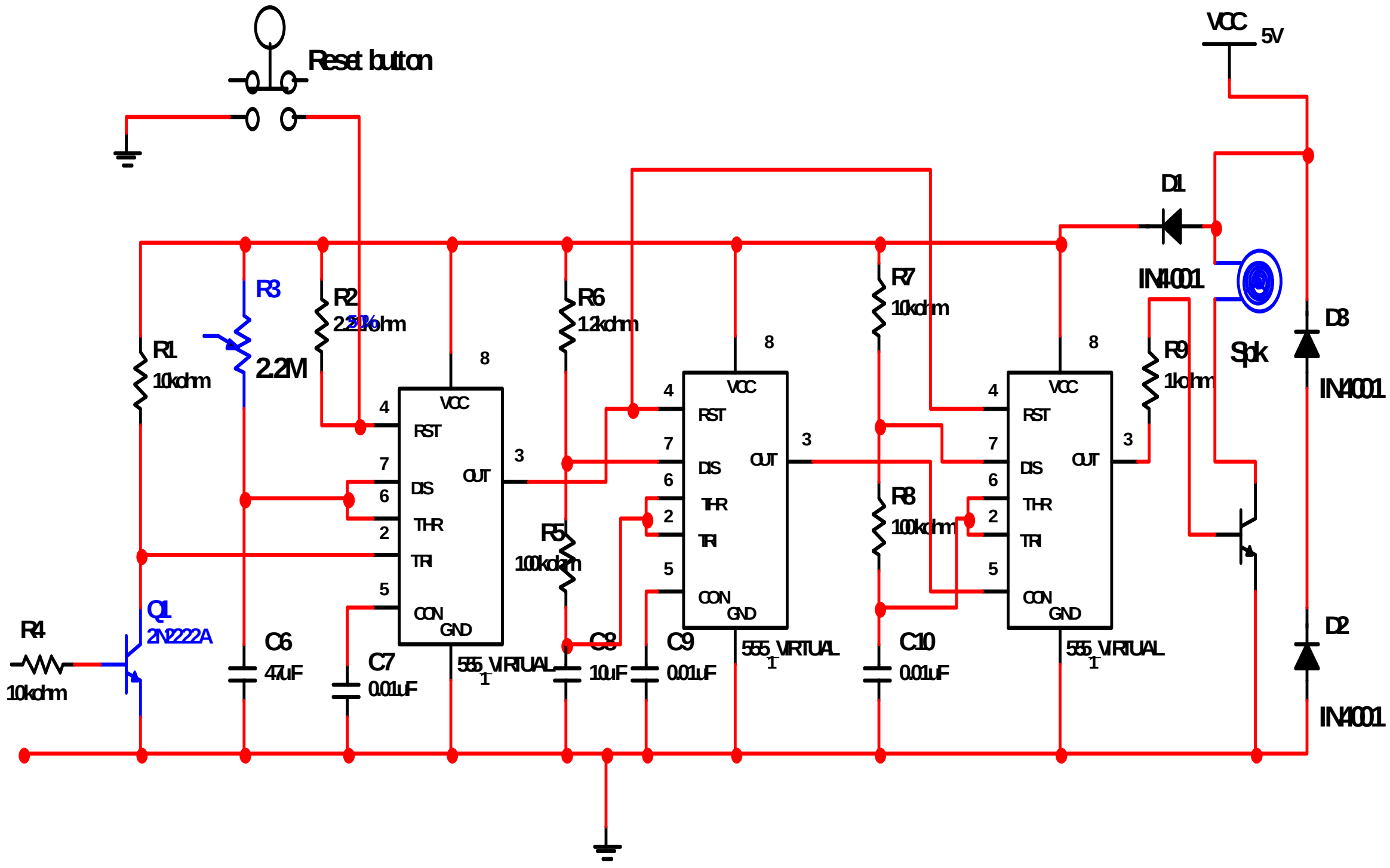

**Fig. 6:** Circuit diagram of the modulated alert system

### 3.4.1. Calculation for the modulated alert tone

For a 555 configured to operate as a monostable oscillator, the time at which it stays high is given as:

$$T = 1.1 \, x \, R \, x \, C \quad (35)$$

A variable resistor (10k$\Omega$) is used for this design work so as to enable us to vary the time at which the alarm stays ON.

Hence for minimum time ($T_2$) = 0 as R = 0

For timer 2 (IC3) it operates in the Astable configuration

$$\text{Hence } T = 0.69\,(R_A + 2R_B)C \quad (36)$$

A time of 1.5 Sec is used so as to give the 500Hz tone a ON and OFF effect

Taking $R_A = 12k\Omega$, $2R_B = \dfrac{1.5}{0.69C} - R_A$

$2R_B = 196333.333$,

Therefore $R_b = 98166.66667 \approx 100k\Omega$

To generate an audible tone of 500Hz

Let $R_A = 10k\Omega$

Therefore $2R_B = 9492.753623 \approx 100k\Omega$

### 3.5. The Power Supply Unit

The power supply unit is subdivided into sections

1. The power supply for the motion detector
2. The power supply for the modulated alert tone

#### 3.5.1. Power supply for the motion detector

This is made up of a 9V battery and 78L05 IC as shown below the 78L05 IC is a voltage regulator which converts the 9V supply to that is needed to drive the detector.

It should be noted that: to avoid changing the battery often we incorporated a DC adaptor pin, so that the device could be hooked up to 9V power supply through the AC to DC adaptor.

#### 3.5.2. Power supply for the modulated alert tone generator

The voltage (9v) needed to drive the alert tone generator is supplied through a 9v battery

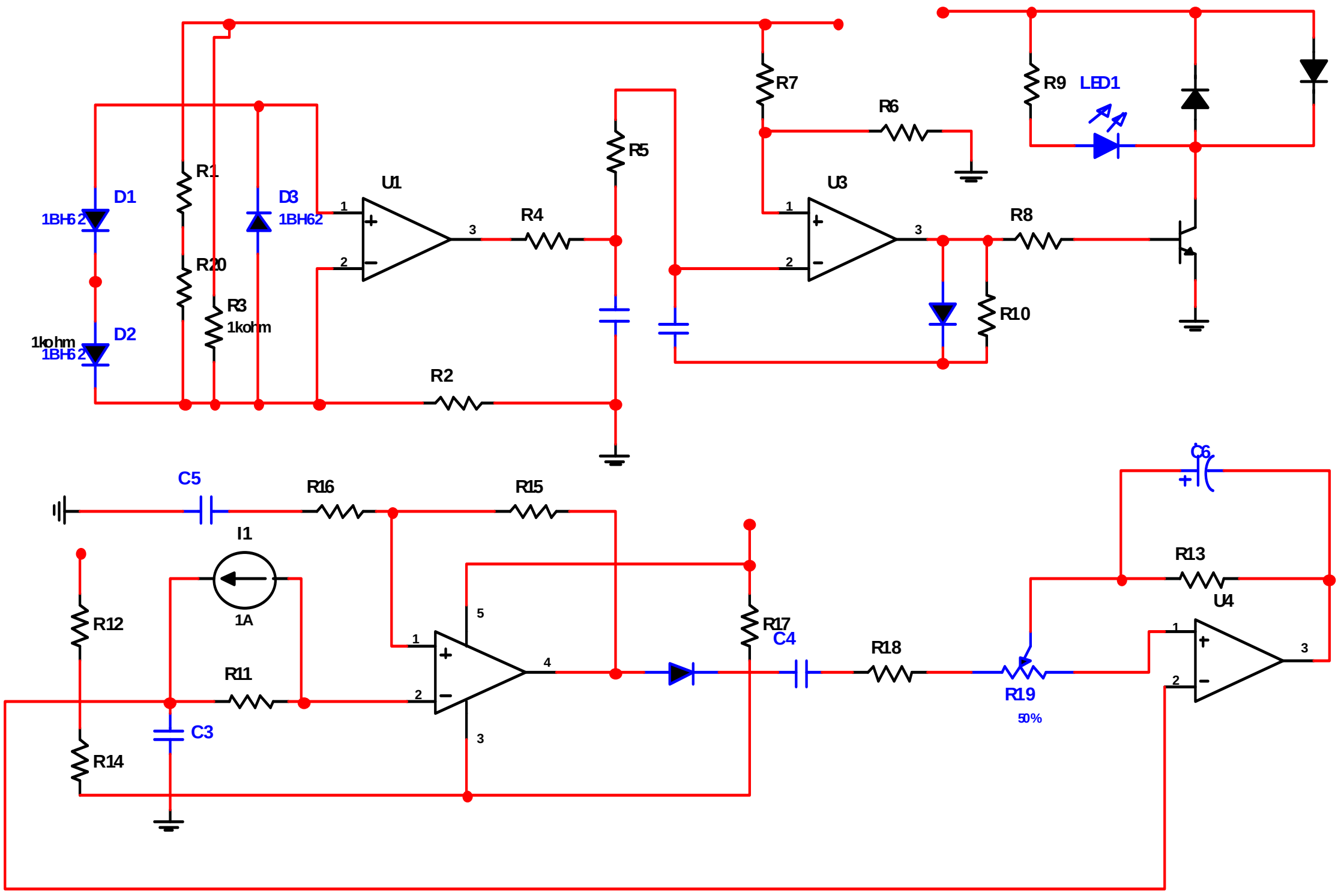

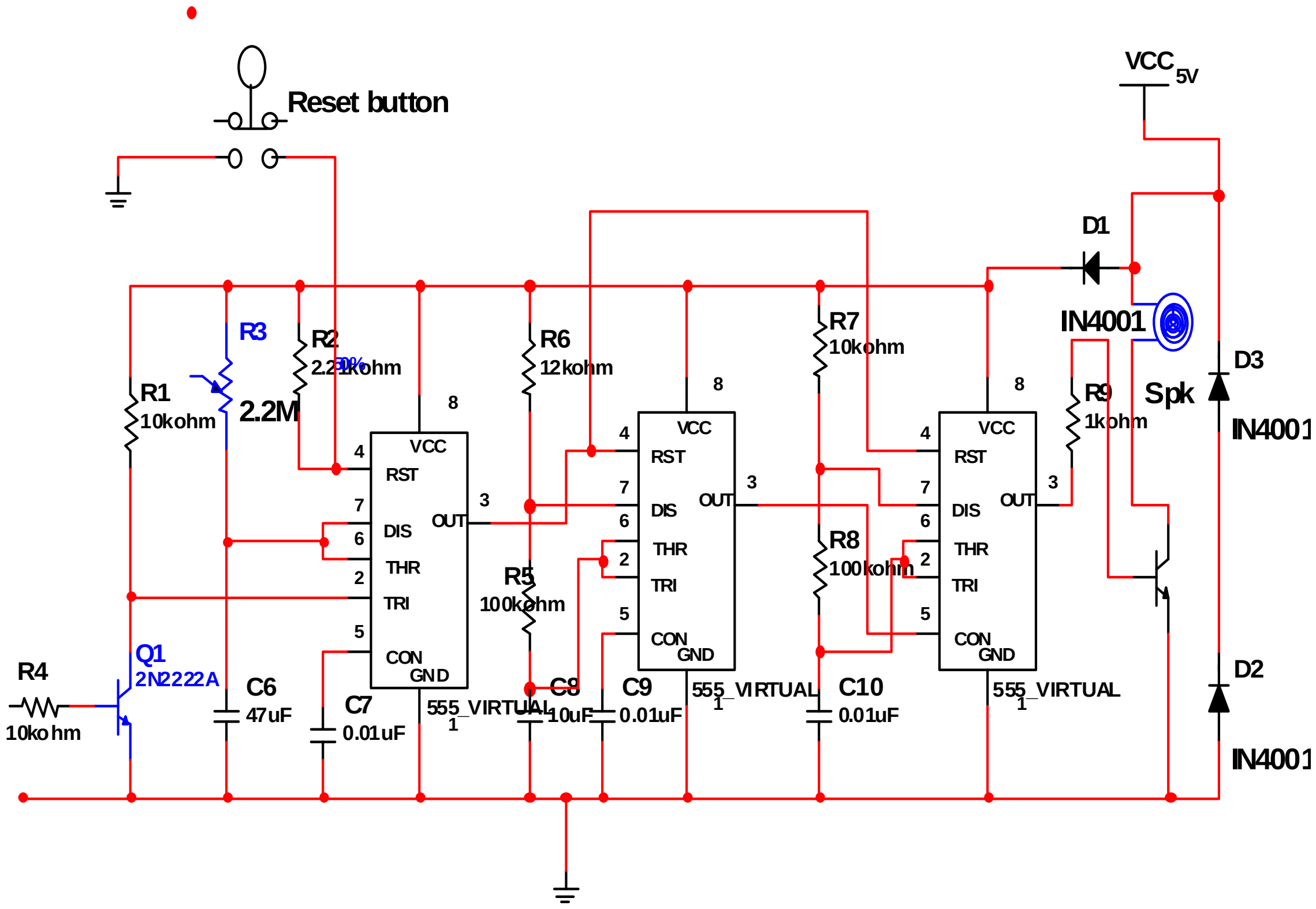

**Fig. 7:** Complete Circuit Diagram of an Ultrasonic Motion Detector

### 3.6. Test and Measurements

After carefully wiring the circuit, checking component polarities and layout of the various segments of the construction were separately tested for their workability. The output of the Astable Oscillator (555 timer) in the transmitter circuit was tested with a frequency meter before connecting the translator to ensure that a frequency of approximately 40KHZ was supplied to the ultrasonic transmitting transducer. The output of the Astable oscillators (555 timer) in the alarm circuit was also tested with a frequency meter to make sure their output signal was in accordance to the required value. The resistors and capacitors were tested to verify their value before the start of construction. After all tested were carried out, the system was found to be very suitable for detecting all the motion within an area of 3-4 meters.

## 4. Conclusions

It could be deduced from the foregoing design analysis that the design of ultrasonic motion detector like any other electronic need careful planning and implementation. There are various motion detectors but this particular one is unique because of its mode of operation, as it transmit exactly 40 kHz waves. This work is mainly a design and construction of a system that has the ability to sense motion through movement of humans or any target, to design a low cost and portable motion detector system, and the design of a system that can be used to trigger another circuit which can trigger ON or OFF the circuit depending on the circuit attached to it. Generally, the design is made to detect movement or moving object in a an enclosed area. In this work, a transmitter transducer generates a signal at a frequency of 40khz, and when the signal is blocked by any moving object, the receiver will be notified and this in turn triggers a buzzer via a timing circuit. This system works on the principle of the signal interference by a moving body. This system works on the principle of the signal interference by a moving body, and the system is dependent on the presence of an intruder or moving object within a monitored area. The system after design and construction was tested and found to work in accordance with specifications.

## ACKNOWLEDGEMENTS

The authors would like to thank Col. Muhammed Sani Bello (RTD), OON, Vice Chairman of MTN Nigeria Communications Limited for supporting the research.

**Authors**

**Engr. (Dr) Adamu Murtala Zungeru** received his BEng in electrical and computer engineering from the Federal University of Technology (FUT) Minna, Nigeria in 2004, MSc in electronic and telecommunication engineering from the Ahmadu Bello University (ABU) Zaria, Nigeria in 2009, and PhD Degree in Electronic and Communication Engineering from the University of Nottingham. He is currently a lecturer two (LII) at the FUT Minna, Nigeria, a position which he started in 2005. He is a registered engineer with the Council for the Regulation of Engineering in Nigeria (COREN), a professional member of the Institute of Electrical and Electronics Engineers (IEEE), and a professional member of the Association for Computing Machinery (ACM). His research interests are in the fields of swarm intelligence, routing algorithms, wireless sensor networks, energy harvesting, automation, home and industrial security system.

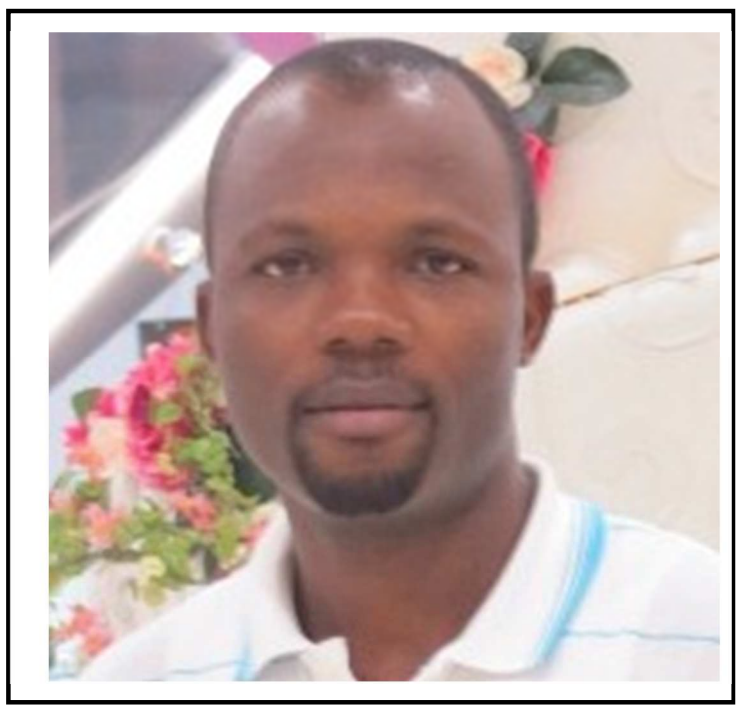